\documentclass[traditabstract]{aa} 
\usepackage{txfonts}
\usepackage{graphicx}
\usepackage{longtable}

\begin{document}

\title{Population of hot subdwarf stars studied with Gaia}
\subtitle{III. Catalogue of known hot subdwarf stars: Data Release 2\thanks{The catalogues are only available in electronic form at the CDS via anonymous ftp to cdsarc.u-strasbg.fr (130.79.128.5) or via http://cdsweb.u-strasbg.fr/cgi-bin/qcat?J/A+A/}}

\author{S.~Geier \inst{1}}

\offprints{S.\,Geier,\\ \email{geier@astro.physik.uni-potsdam.de}}

\institute{Institut f\"ur Physik und Astronomie, Universit\"at Potsdam, Haus 28, Karl-Liebknecht-Str. 24/25, D-14476 Potsdam-Golm, Germany}

\date{Received \ Accepted}

\abstract{In light of substantial new discoveries of hot subdwarfs by ongoing spectroscopic surveys and the availability of new all-sky data from ground-based photometric surveys and the {\em Gaia} mission Data Release 2, we compiled an updated catalogue of the known hot subdwarf stars. The catalogue contains 5874 unique sources including 528 previously unknown hot subdwarfs and provides multi-band photometry, astrometry from {\em Gaia,} and classifications based on spectroscopy and colours. This new catalogue provides atmospheric parameters of 2187 stars and radial velocities of 2790 stars from the literature. Using colour, absolute magnitude, and reduced proper motion criteria, we identified 268 previously misclassified objects, most of which are less luminous white dwarfs or more luminous blue horizontal branch and main-sequence stars.} 


\maketitle

\section{Introduction \label{sec:intro}}

Hot subdwarf stars (sdO/Bs) are situated at the extreme blue end of the horizontal branch (HB), the extreme horizontal branch (EHB; Heber et al. \cite{heber86}). To evolve to the EHB, red giants must lose almost their entire hydrogen envelopes. This is best explained by various scenarios of binary mass transfer (Han et al. \cite{han02,han03}; see Heber \cite{heber16} for a review). Although not initially recognised as such, sdO/B stars were discovered via photometric surveys of faint blue stars (Humason \& Zwicky \cite{humason47}; Iriarte \& Chavira \cite{iriarte57}; Chavira \cite{chavira58,chavira59}; Haro \& Luyten \cite{haro62}; Green et al. \cite{green86}; Downes \cite{downes86}). Subsequently, Kilkenny et al. (\cite{kilkenny88}) published the first catalogue of 1225 spectroscopically identified hot subdwarf stars. Objective prism surveys obtaining low-resolution spectra detected many more hot subdwarfs (Hagen et al. \cite{hagen95}; Wisotzki et al. \cite{wisotzki96}; Stobie et al. \cite{stobie97}; Mickaelian et al. \cite{mickaelian07,mickaelian08}). \O stensen (\cite{oestensen06}) compiled a database containing more than 2300 stars. 

Subsequently, the Sloan Digital Sky Survey (SDSS) provided spectra of almost 2000 sdO/Bs (Geier et al. \cite{geier15b}; Kepler et al. \cite{kepler15,kepler16}) and new samples of bright hot subdwarf stars were selected (e.g. Vennes et al. \cite{vennes11}). Furthermore, data from new large-area photometric and astrometric surveys have been conducted in multiple bands from the UV to the far-infrared. 

This motivated us to compile a new catalogue of hot subdwarf stars (Geier et al. \cite{geier17a}). We started with the catalogue of \O stensen (\cite{oestensen06}) and added hot subdwarf candidates from several recent spectroscopic surveys (Mickaelian et al. \cite{mickaelian08}; \O stensen et al. \cite{oestensen10b}; Geier et al. \cite{geier15b}; Kepler et al. \cite{kepler16}; Gentile Fusillo et al. (\cite{gentile15}; O'Donoghue et al. \cite{odonoghue13}; Kilkenny et al. \cite{kilkenny15,kilkenny16}; Vennes et al. (\cite{vennes11}; Oreiro et al. (\cite{oreiro11}; Perez-Fernandez et al. (\cite{perez16}; Luo et al. \cite{luo16}) and unpublished sources. We cross-matched all those objects with large-area photometric and light curve survey catalogues. Proper motions were obtained from diverse ground-based surveys (see Geier et al. \cite{geier17a} for details). 

The spectroscopic catalogue of Geier et al. (\cite{geier17a}) has been used as an input catalogue for photometric (TESS; Stassun et al. \cite{stassun18}) and spectroscopic surveys (LAMOST; Lei et al. \cite{lei20}); it has also been used to select stars for more detailed studies (Boudreaux et al. \cite{boudreaux17}; Carrillo et al. \cite{carrillo20}). Furthermore, this previous catalogue has been used to determine selection criteria for an all-sky catalogue of hot subluminous star candidates selected from {\em Gaia} Data Release 2 (DR2; {\em Gaia} Collaboration et al. \cite{gaia18}) by means of colour, absolute magnitude, and reduced proper motion cuts (Geier et al. \cite{geier19}). In this work, I present Data Release 2 of the catalogue of known hot subdwarf stars.

\begin{figure}[t!]
\begin{center}
        \resizebox{9cm}{!}{\includegraphics{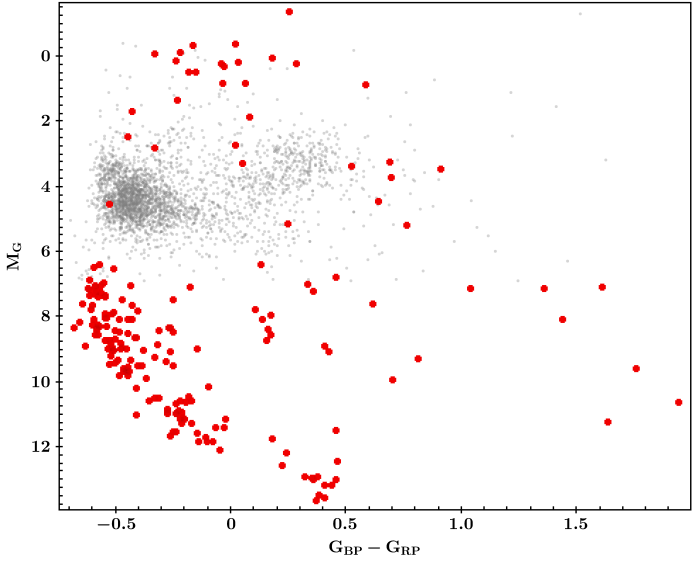}}
\end{center} 
\caption{Gaia colour-absolute magnitude diagram of the subsample with accurate parallaxes. Grey dots denote sdO/Bs and red dots show misclassified objects.}
\label{missclass}
\end{figure}

\begin{table}
\caption{\label{tab2} Colour-classification schemes}
\begin{center}
\begin{tabular}{ll}
\hline\hline
\noalign{\smallskip}
SDSS & \\
\noalign{\smallskip}
\hline
\noalign{\smallskip}
sdO & $-0.55<u_{\rm SDSS}-g_{\rm SDSS}<-0.35$ \\
    & $-0.65<g_{\rm SDSS}-r_{\rm SDSS}<-0.45$ \\
sdB & $-0.5<u_{\rm SDSS}-g_{\rm SDSS}<0.7$ \\
    & $g_{\rm SDSS}-r_{\rm SDSS}>0.208(u_{\rm SDSS}-g_{\rm SDSS})-0.516$ \\
    & $g_{\rm SDSS}-r_{\rm SDSS}<0.208(u_{\rm SDSS}-g_{\rm SDSS})-0.376$ \\
sd+MS & $-0.5<u_{\rm SDSS}-g_{\rm SDSS}<0.7$ \\
      & $g_{\rm SDSS}-r_{\rm SDSS}>0.208(u_{\rm SDSS}-g_{\rm SDSS})-0.376$ \\
\noalign{\smallskip}
\hline
\noalign{\smallskip}
GALEX/APASS & \\
\noalign{\smallskip}
\hline
\noalign{\smallskip}
sdO/B & $NUV_{\rm GALEX}-g_{\rm APASS}<2.0$ \\
      & $g_{\rm APASS}-r_{\rm APASS}<-0.15$ \\
sd+MS & $NUV_{\rm GALEX}-g_{\rm APASS}<2.0$ \\
      & $g_{\rm APASS}-r_{\rm APASS}\geq-0.15$ \\
\noalign{\smallskip}
\hline
\noalign{\smallskip}
GALEX/PS1 & \\
\noalign{\smallskip}
\hline
\noalign{\smallskip}
sdO/B & $NUV_{\rm GALEX}-g_{\rm PS1}<1.7$ \\
      & $g_{\rm PS1}-r_{\rm PS1}<-0.2$ \\
sd+MS & $NUV_{\rm GALEX}-g_{\rm PS1}<1.7$ \\
      & $g_{\rm PS1}-r_{\rm APASS}\geq-0.2$ \\
\noalign{\smallskip}
\hline
\noalign{\smallskip}
SkyMapper & \\
\noalign{\smallskip}
\hline
\noalign{\smallskip}
sdO & $-0.8<u_{\rm SKYM}-g_{\rm SKYM}<-0.4$ \\
    & $-0.5<g_{\rm SKYM}-r_{\rm SKYM}<-0.17$ \\
sdB & $-0.4<u_{\rm SKYM}-g_{\rm SKYM}<0.9$ \\
    & $-0.5<g_{\rm SKYM}-r_{\rm SKYM}<-0.17$ \\
sd+MS & $-0.8<u_{\rm SKYM}-g_{\rm SKYM}<0.9$ \\
    & $-0.17<g_{\rm SKYM}-r_{\rm SKYM}<0.3$ \\
BHB & $0.9<u_{\rm SKYM}-g_{\rm SKYM}<1.4$ \\
    & $0.1<g_{\rm SKYM}-r_{\rm SKYM}<0.55$ \\    
\noalign{\smallskip}
\hline\hline
\end{tabular}
\end{center}
\end{table} 

\begin{figure*}[t!]
\begin{center}
        \resizebox{15cm}{!}{\includegraphics{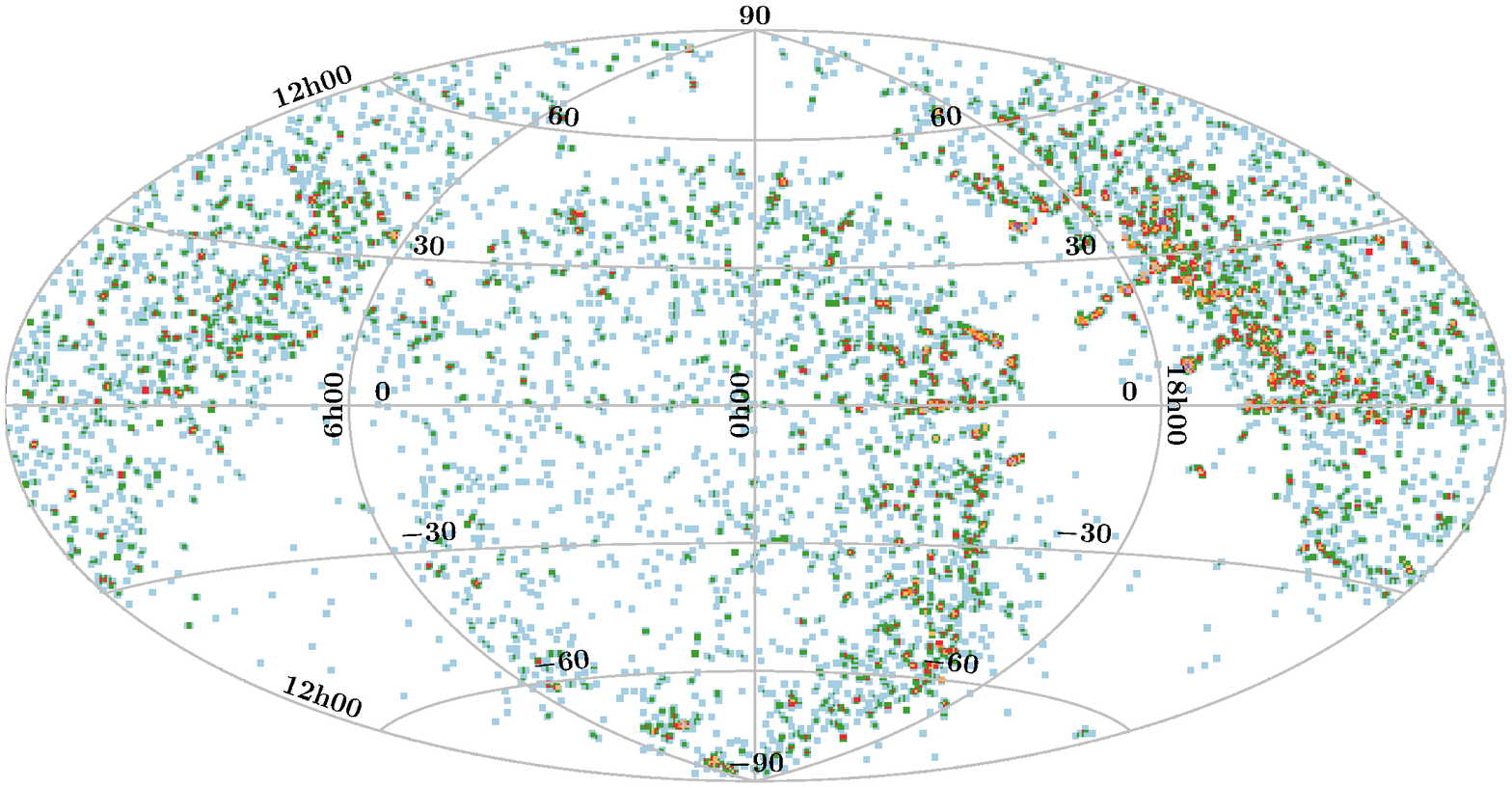}}
\end{center} 
\caption{\bf Sky distribution of the sdO/Bs catalogue in equatorial coordinates. The colour scale represents the densities of the stars starting from green to red, and yellow indicates the most crowded regions.}
\label{coordinates}
\end{figure*}

\section{Constructing the catalogue Data Release 2} 

\subsection{Input data}

In addition to the spectroscopically classified hot subdwarf stars from Data Release 1 (DR1) of this catalogue (Geier et al. \cite{geier17a}), several new samples of hot subdwarfs have recently been published, most of which have atmospheric parameters and radial velocity determinations. Kepler et al. (\cite{kepler19}) identified sdO/Bs in SDSS DR14 and Geier et al. \cite{geier17b} provided atmospheric parameters for a large sample of hot subdwarfs from SDSS DR7. However, the most important new source of yet undiscovered sdO/Bs is the LAMOST survey (Lei et al. \cite{lei18,lei19,lei20}; Luo et al. \cite{luo19}). It has to be pointed out that a lot of the recently published stars had already been classified and that there are large overlaps between the samples. By carefully cross-matching the new samples with the catalogue and each other, I found that 528 new sdO/Bs have been discovered since the publication of DR1. 

New hot subdwarfs have also been identified in globular clusters (Latour et al. \cite{latour18}), but owing to the different types of photometric and astrometric data available for those objects, they are not included in this catalogue. This means that the catalogue should be regarded as compilation of known hot subdwarf stars in the field.

\subsection{Multi-band photometry}

Near-ultraviolet (NUV) and far-ultraviolet (FUV) photometry were taken from the Galaxy Evolution Explorer (GALEX) All-sky Imaging Survey (AIS DR5; Bianchi et al. \cite{bianchi11}). Optical photometry was obtained from Gaia DR2 in the $G_{\rm BP},G_{\rm RP}$, and $G$ bands ({\em Gaia} Collaboration et al. \cite{gaia18}), the American Association of Variable Star Observers (AAVSO) Photometric All Sky Survey (APASS DR9; Henden et al. \cite{henden16}) in the $VBgri$ bands, the SDSS DR12 (Alam et al. \cite{alam15}) in the $ugriz$ bands, the Panoramic Survey Telescope and Rapid Response System (Pan-STARRS) PS1 survey (Chambers et al. \cite{chambers16}) in the $grizy$ bands, the SkyMapper Southern Sky Survey DR1.1 (Wolf et al. \cite{wolf18}) in the $uvgriz$ bands, and the Very Large Telescope Survey Telescope (VST) ATLAS (DR3; Shanks et al. \cite{shanks15}) and the Kilo-Degree (KiDS DR3; de Jong et al. \cite{dejong15}) ESO public surveys in the $ugriz$ bands. 

Near-infrared photometry was obtained from the Two Micron All-sky Survey (2MASS) All-Sky catalogue of Point Sources (Skrutskie et al. \cite{skrutskie06}) in the $JHK$ bands, the United Kingdom Infra-Red Telescope (UKIRT) Infrared Deep Sky Survey (UKIDSS Large Area Survey DR9; Lawrence et al. \cite{lawrence07}) in the $YJHK$ bands, the Visible and Infrared Survey Telescope for Astronomy (VISTA) Hemisphere (VHS DR2; McMahon et al. \cite{mcmahon13}), and the VISTA Kilo-degree Infrared Galaxy (VIKING DR4; Edge et al. \cite{edge13}) ESO public surveys in the $ZYJHK_{\rm S}$ bands. Far-infrared photometry was obtained from the Wide-field Infrared Survey Explorer mission (WISE) AllWISE data release (Cutri et al. \cite{cutri14}) in the four WISE bands. Galactic reddening $E(B-V)$ and  Galactic dust extinction $A_V$ from the maps of Schlafly \& Finkbeiner (\cite{schlafly11}) were also provided. 

\subsection{Gaia astrometry}

{\em Gaia} DR2 ({\em Gaia} Collaboration et al. \cite{gaia18}) provides precise coordinates, parallaxes, and proper motions (Lindegren et al. \cite{lindegren18}), which have been included in the catalogue. Ground-based proper motions are no longer included.   

\subsection{Cleaning the catalogue}\label{cleaning}

The data collected were used to identify and remove objects misclassified as hot subdwarf stars (Geier et al. \cite{geier17a}). To separate all kinds of cooler objects, colour indices were used. Objects with SDSS colours $u-g>0.6$ and $g-r>0.1$, $NUV_{\rm GALEX}-g_{\rm PS1}>1.7$, SkyMapper $u-g>0.9$ and $NUV_{\rm GALEX}-g_{\rm APASS}>2.0$ have been excluded if different indices were consistent. If only one of those indices was available, Gaia $G_{\rm BP}-G_{\rm RP}>0.4$ was used as additional constraint. 

{\em Gaia} DR2 provides us with accurate parallax distances with uncertainties smaller than $20\%$ for distances up to about $1-2\,{\rm kpc}$ (Lindegren et al. \cite{lindegren18}). Calculating the absolute magnitudes of all stars with accurate parallaxes, it was possible to identify misclassified white dwarfs (WDs) and also brighter objects such as blue horizontal branch (BHB) or main-sequence B  (MS-B) stars (see Fig,~\ref{missclass}). 

To distinguish WDs from the more luminous and distant sdO/Bs, I also used the reduced proper motion method as outlined in Gentile Fusillo et al. (\cite{gentile15}). The reduced proper motion $H=G+5\log{\mu }+5$ was calculated using the {\em Gaia} G magnitudes and proper motions because accurate {\em Gaia} proper motions are available for all stars in the catalogue. Stars with $H>15$ are WD candidates and therefore excluded (Geier et al. \cite{geier17a}). Most of the WD candidates identified in this work are also listed in the Gaia WD catalogue (Gentile Fusillo et al. \cite{gentile19}). Some bright BHB and MS-B stars have been identified from new follow-up spectra (Schneider et al. in prep.). The catalogue was also cross-matched with SIMBAD and misclassified objects known from the literature were also excluded.

The 268 misclassified objects are provided with their correct classifications as a separate catalogue. Based on their previous classifications as sdBs or sdOs and their colours, a tentative classification of the WD candidates as either DAB (hydrogen and possibly neutral helium lines) or DAO (hydrogen and/or ionised helium lines) candidates is provided. If there was no detailed classification before, they are classified as WD.  
 
The final DR2 catalogue contains 5874 unique objects (see Fig.~\ref{coordinates}). Thanks to {\em Gaia}, which allows us to separate the WDs and bright MS-B stars very efficiently, BHB stars should now be the most important class of contaminant objects remaining in the catalogue.

\subsection{Classification of hot subdwarfs}

For spectroscopic and photometric classifications, we follow the scheme outlined in Geier et al. (\cite{geier17a}). The empirical scheme for the photometric classification by inspecting the locations of the subclasses in two-colour diagrams could be extended using new colour criteria based on multi-band photometry from PanSTARRS (Chambers et al. \cite{chambers16}) and SkyMapper (Wolf et al. \cite{wolf18}). 

\begin{figure}[t!]
\begin{center}
        \resizebox{9cm}{!}{\includegraphics{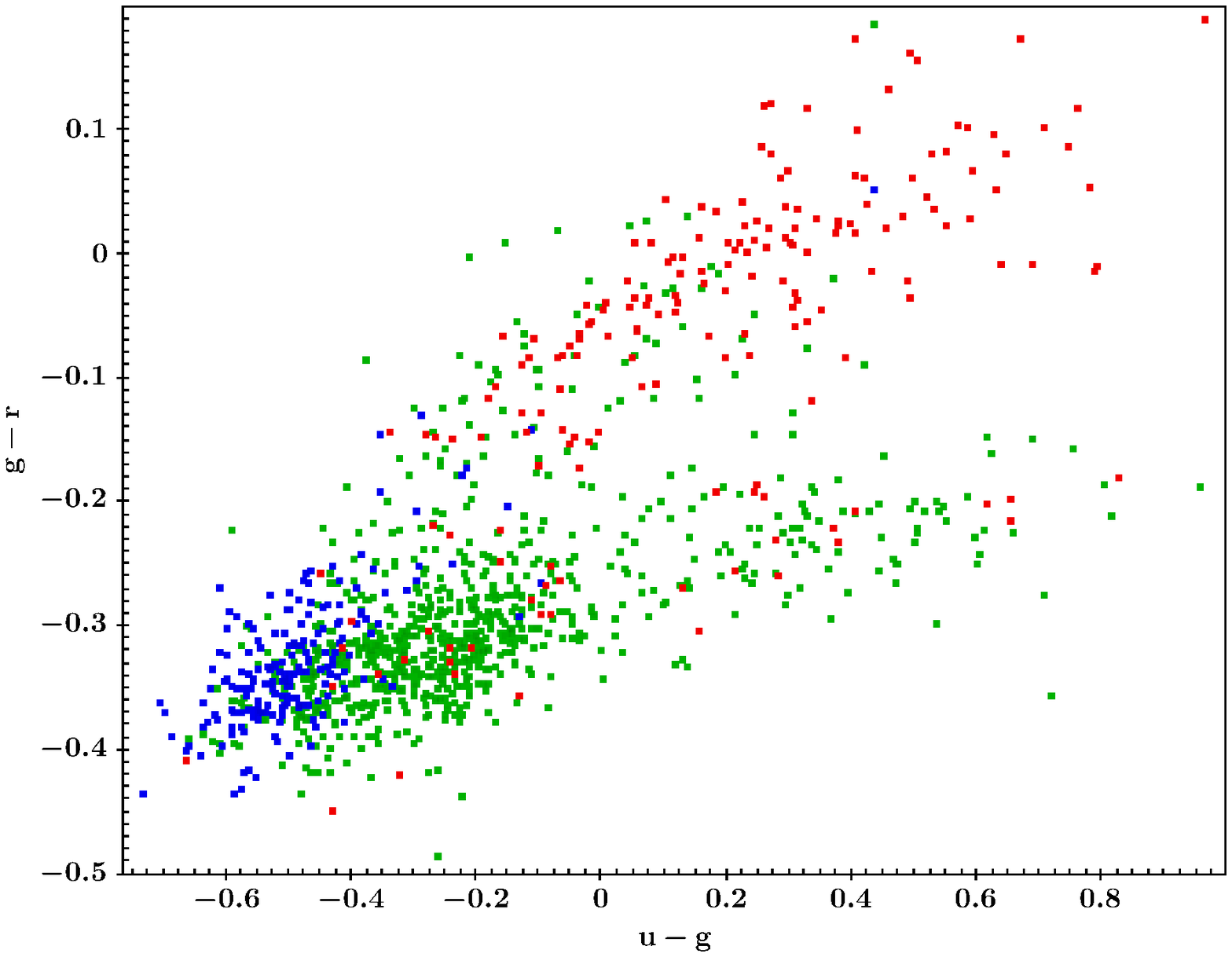}}
        \resizebox{9cm}{!}{\includegraphics{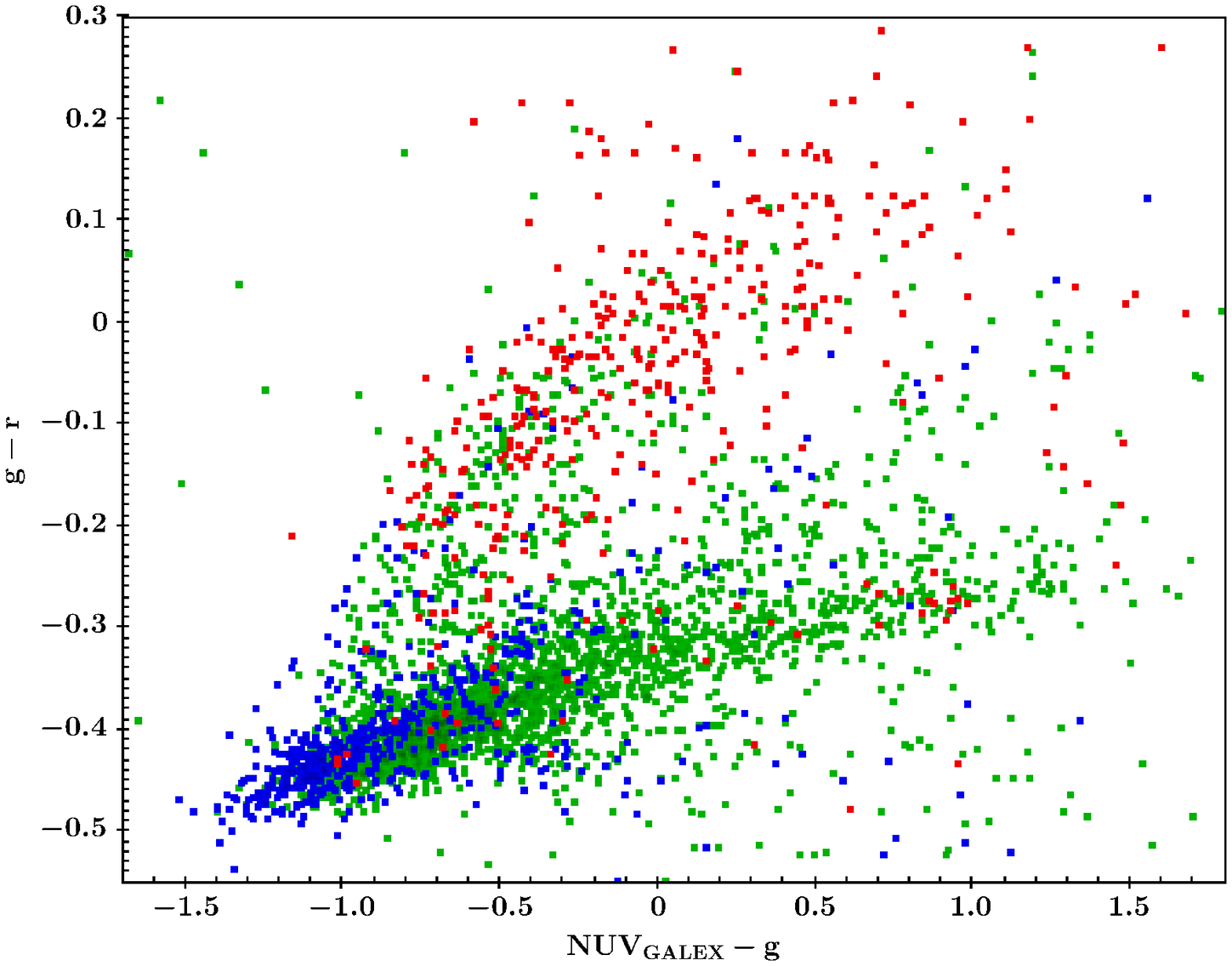}}
\end{center} 
\caption{Two-colour diagrams for spectroscopically classified objects from the hot subdwarf catalogue. The sdB and sdOB stars are indicated in green, sdO stars in blue, and composite binaries in red. Upper panel: Skymapper. Lower panel: GALEX/PS1.}
\label{colour}
\end{figure}

The SDSS- and SkyMapper-based colour classes should be regarded as the most trustworthy because the u-g colour allows us to distinguish between sdB and the hotter sdO types better than a combination of NUV and g band (see Fig.~\ref{colour}). The updated colour criteria are provided in Table~\ref{tab2}. 

\subsection{Spectroscopic parameters and radial velocities}

The catalogue contains spectroscopic parameters such as effective temperatures, surface gravities, and helium abundances for 2187 stars from the literature; this is  more than twice as many as in DR1. This fraction is still not complete because only papers containing larger samples of sdO/B stars have been taken into account (Heber et al. \cite{heber84}; Bixler et al. \cite{bixler91}; Saffer et al. \cite{saffer94,saffer97}; Maxted et al. \cite{maxted01}; Edelmann et al. \cite{edelmann03}; Lisker et al. \cite{lisker05}; Str\"oer et al. \cite{stroeer07}; Hirsch \cite{hirsch09}; \O stensen et al. \cite{oestensen10a}; Nemeth et al. \cite{nemeth12}; Geier et al. \cite{geier13,geier15b,geier17b}; Kupfer et al. \cite{kupfer15}; Luo et al. \cite{luo16,luo19}; Kepler et al. \cite{kepler16,kepler19}; Lei et al. \cite{lei18,lei19,lei20}). Since the main purpose of this catalogue is the identification and classification of hot subdwarf stars, only one set of atmospheric parameters, which is regarded as reliable, is provided for each star even if several different determinations are provided in the literature. 

Radial velocities (RVs) are provided for the 2790 stars with spectra in the SDSS and LAMOST data archives. However, for the most helium rich objects, systematic offsets by up to $\sim100\,{\rm km\,s^{-1}}$ are possible owing to cross-correlation with inadequate template spectra (Geier et al. \cite{geier15a,geier17a}). 

\section{Summary}

The catalogue of known hot subdwarf stars DR2 and the catalogue of objects previously misclassified as hot subdwarfs are both available via the VizieR service. A detailed description of the catalogue columns for both catalogues is provided in Table~\ref{tab3}. The catalogue is by no means complete and heterogeneously selected, which has to be taken into account when using it for statistical analyses. The large samples from LAMOST for example only include single-lined sdO/Bs and the composite sdB+MS binaries are supposed to be published in the future. 

DR2 contains 528 newly discovered sdO/B stars and is significantly cleaner than DR1. With the growing number of large-area spectroscopic surveys it is important to keep track of the known objects to allow for a most efficient follow-up of the yet unclassified candidates. The multitude of new designations that comes with each survey does not make this effort easier. 

The previously missclassified objects are a mixed bag of interesting and often peculiar stars. The MS-B stars for example are all quite faint and found mostly at high Galactic latitudes. This indicates large distances and a likely runaway origin. Most WDs are likely quite hot, young, and short-lived, and therefore rare. 

The Gaia catalogue of hot subluminous star candidates will be maintained and updated in parallel (Geier et al. \cite{geier19}) to provide a comprehensive list of new hot subdwarf candidates. Together with the WD catalogue (Gentile Fusillo et al. \cite{gentile19}) and the catalogue of extremely low-mass WD candidates (Pelisoli \& Vos \cite{pelisoli19}) the whole parameter space of hot subluminous stars is now covered. In addition, we are currently compiling catalogues of variable hot subdwarfs based on the diverse light-curve surveys now available. The first such catalogue including eclipsing HW\,Vir-type binaries has recently been published by Schaffenroth et al. (\cite{schaffenroth19}).

\begin{acknowledgements}

I would like to thank the referee Dave Kilkenny for his constructive report. 

S.G. was supported by the Heisenberg program of the Deutsche Forschungsgemeinschaft (DFG) through grants GE 2506/8-1 and GE 2506/9-1. 

This research made use of TOPCAT, an interactive graphical viewer and editor for tabular data Taylor (\cite{taylor05}). This research made use of the SIMBAD database, operated at CDS, Strasbourg, France; the VizieR catalogue access tool, CDS, Strasbourg, France. Some of the data presented in this paper were obtained from the Mikulski Archive for Space Telescopes (MAST). STScI is operated by the Association of Universities for Research in Astronomy, Inc., under NASA contract NAS5-26555. Support for MAST for non-HST data is provided by the NASA Office of Space Science via grant NNX13AC07G and by other grants and contracts. This research has made use of the services of the ESO Science Archive Facility.

This work has made use of data from the European Space Agency (ESA) mission {\it Gaia} (https://www.cosmos.esa.int/gaia), processed by the {\it Gaia} Data Processing and Analysis Consortium (DPAC, https://www.cosmos.esa.int/web/gaia/dpac/consortium). Funding for the DPAC has been provided by national institutions, in particular the institutions participating in the {\it Gaia} Multilateral Agreement.

This publication makes use of data products from the Two Micron All Sky Survey, which is a joint project of the University of Massachusetts and the Infrared Processing and Analysis Center/California Institute of Technology, funded by the National Aeronautics and Space Administration and the National Science Foundation. Based on observations made with the NASA Galaxy Evolution Explorer. GALEX is operated for NASA by the California Institute of Technology under NASA contract NAS5-98034. This research has made use of the APASS database, located at the AAVSO web site. Funding for APASS has been provided by the Robert Martin Ayers Sciences Fund. The Guide Star catalogue-II is a joint project of the Space Telescope Science Institute and the Osservatorio Astronomico di Torino. Space Telescope Science Institute is operated by the Association of Universities for Research in Astronomy, for the National Aeronautics and Space Administration under contract NAS5-26555. The participation of the Osservatorio Astronomico di Torino is supported by the Italian Council for Research in Astronomy. Additional support is provided by European Southern Observatory, Space Telescope European Coordinating Facility, the International GEMINI project and the European Space Agency Astrophysics Division.

Based on observations obtained as part of the VISTA Hemisphere Survey, ESO Program, 179.A-2010 (PI: McMahon). This publication has made use of data from the VIKING survey from VISTA at the ESO Paranal Observatory, programme ID 179.A-2004. Data processing has been contributed by the VISTA Data Flow System at CASU, Cambridge and WFAU, Edinburgh. Based on data products from observations made with ESO Telescopes at the La Silla Paranal Observatory under program ID 177.A 3011(A,B,C,D,E.F). Based on data products from observations made with ESO Telescopes at the La Silla Paranal Observatory under programme IDs 177.A-3016, 177.A-3017 and 177.A-3018, and on data products produced by Target/OmegaCEN, INAF-OACN, INAF-OAPD and the KiDS production team, on behalf of the KiDS consortium. OmegaCEN and the KiDS production team acknowledge support by NOVA and NWO-M grants. Members of INAF-OAPD and INAF-OACN also acknowledge the support from the Department of Physics \& Astronomy of the University of Padova, and of the Department of Physics of Univ. Federico II (Naples). This publication makes use of data products from the Wide-field Infrared Survey Explorer, which is a joint project of the University of California, Los Angeles, and the Jet Propulsion Laboratory/California Institute of Technology, and NEOWISE, which is a project of the Jet Propulsion Laboratory/California Institute of Technology. WISE and NEOWISE are funded by the National Aeronautics and Space Administration.

Based on observations collected at the Centro Astron\'omico Hispano Alem\'an (CAHA) at Calar Alto, operated jointly by the Max-Planck Institut f\"ur Astronomie and the Instituto de Astrof\'isica de Andaluc\'ia (CSIC). Based on observations with the William Herschel and Isaac Newton Telescopes operated by the Isaac Newton Group at the Observatorio del Roque de los Muchachos of the Instituto de Astrofisica de Canarias on the island of La Palma, Spain. 

The Pan-STARRS1 Surveys (PS1) and the PS1 public science archive have been made possible through contributions by the Institute for Astronomy, the University of Hawaii, the Pan-STARRS Project Office, the Max-Planck Society and its participating institutes, the Max Planck Institute for Astronomy, Heidelberg and the Max Planck Institute for Extraterrestrial Physics, Garching, The Johns Hopkins University, Durham University, the University of Edinburgh, the Queen's University Belfast, the Harvard-Smithsonian Center for Astrophysics, the Las Cumbres Observatory Global Telescope Network Incorporated, the National Central University of Taiwan, the Space Telescope Science Institute, the National Aeronautics and Space Administration under Grant No. NNX08AR22G issued through the Planetary Science Division of the NASA Science Mission Directorate, the National Science Foundation Grant No. AST-1238877, the University of Maryland, Eotvos Lorand University (ELTE), the Los Alamos National Laboratory, and the Gordon and Betty Moore Foundation.

The national facility capability for SkyMapper has been funded through ARC LIEF grant LE130100104 from the Australian Research Council, awarded to the University of Sydney, the Australian National University, Swinburne University of Technology, the University of Queensland, the University of Western Australia, the University of Melbourne, Curtin University of Technology, Monash University and the Australian Astronomical Observatory. SkyMapper is owned and operated by The Australian National University's Research School of Astronomy and Astrophysics. The survey data were processed and provided by the SkyMapper Team at ANU. The SkyMapper node of the All-Sky Virtual Observatory (ASVO) is hosted at the National Computational Infrastructure (NCI). Development and support the SkyMapper node of the ASVO has been funded in part by Astronomy Australia Limited (AAL) and the Australian Government through the Commonwealth's Education Investment Fund (EIF) and National Collaborative Research Infrastructure Strategy (NCRIS), particularly the National eResearch Collaboration Tools and Resources (NeCTAR) and the Australian National Data Service Projects (ANDS).

Guoshoujing Telescope (the Large Sky Area Multi-Object Fiber Spectroscopic Telescope LAMOST) is a National Major Scientific Project built by the Chinese Academy of Sciences. Funding for the project has been provided by the National Development and Reform Commission. LAMOST is operated and managed by the National Astronomical Observatories, Chinese Academy of Sciences.

Funding for the SDSS and SDSS-II has been provided by the Alfred P. Sloan Foundation, the Participating Institutions, the National Science Foundation, the U.S. Department of Energy, the National Aeronautics and Space Administration, the Japanese Monbukagakusho, the Max Planck Society, and the Higher Education Funding Council for England. The SDSS Web Site is http://www.sdss.org/. The SDSS is managed by the Astrophysical Research Consortium for the Participating Institutions. The Participating Institutions are the American Museum of Natural History, Astrophysical Institute Potsdam, University of Basel, University of Cambridge, Case Western Reserve University, University of Chicago, Drexel University, Fermilab, the Institute for Advanced Study, the Japan Participation Group, Johns Hopkins University, the Joint Institute for Nuclear Astrophysics, the Kavli Institute for Particle Astrophysics and Cosmology, the Korean Scientist Group, the Chinese Academy of Sciences (LAMOST), Los Alamos National Laboratory, the Max-Planck-Institute for Astronomy (MPIA), the Max-Planck-Institute for Astrophysics (MPA), New Mexico State University, Ohio State University, University of Pittsburgh, University of Portsmouth, Princeton University, the United States Naval Observatory, and the University of Washington. 

Funding for SDSS-III has been provided by the Alfred P. Sloan Foundation, the Participating Institutions, the National Science Foundation, and the U.S. Department of Energy Office of Science. The SDSS-III web site is http://www.sdss3.org/. SDSS-III is managed by the Astrophysical Research Consortium for the Participating Institutions of the SDSS-III Collaboration including the University of Arizona, the Brazilian Participation Group, Brookhaven National Laboratory, University of Cambridge, Carnegie Mellon University, University of Florida, the French Participation Group, the German Participation Group, Harvard University, the Instituto de Astrofisica de Canarias, the Michigan State/Notre Dame/JINA Participation Group, Johns Hopkins University, Lawrence Berkeley National Laboratory, Max Planck Institute for Astrophysics, Max Planck Institute for Extraterrestrial Physics, New Mexico State University, New York University, Ohio State University, Pennsylvania State University, University of Portsmouth, Princeton University, the Spanish Participation Group, University of Tokyo, University of Utah, Vanderbilt University, University of Virginia, University of Washington, and Yale University. 

\end{acknowledgements}

\clearpage
\onecolumn

\begin{longtable}{llll}
\caption{\label{tab3} Catalogue columns}\\
\hline\hline
\noalign{\smallskip}
Column & Format & Description & Unit \\
\noalign{\smallskip}
\hline
\noalign{\smallskip}
NAME & A30 & Target name & \\
GAIA\_DESIG & A30 & Gaia designation & \\
RA & F10.6 & Right ascension (J2000) & deg \\
DEC & F10.6 & Declination (J2000) & deg \\
GLON & F10.6 & Galactic longitude & deg \\
GLAT & F10.6 & Galactic latitude & deg \\
SPEC\_CLASS & A15 & Spectroscopic classification & \\
SPEC\_SIMBAD & A15 & Spectroscopic classification from SIMBAD & \\
COLOUR\_SDSS & A10 & Colour classification SDSS & \\
COLOUR\_APASS & A10 & Colour classification GALEX/APASS & \\
COLOUR\_PS1 & A10 & Colour classification GALEX/PS1 & \\
COLOUR\_SKYM & A10 & Colour classification SkyMapper & \\
PLX & F8.4 & Gaia parallax & mas \\
e\_PLX & F8.4 & Error on PLX & mas \\
M\_G   & F8.4 & Absolute magnitude in G-band & mag \\
G\_GAIA & F6.3 & Gaia G-band magnitude & mag \\
e\_G\_GAIA & F6.3 & Error on G\_GAIA & mag \\
BP\_GAIA & F6.3 & Gaia BP-band magnitude & mag \\
e\_BP\_GAIA & F6.3 & Error on BP\_GAIA & mag \\
RP\_GAIA & F6.3 & Gaia RP-band magnitude & mag \\
e\_RP\_GAIA & F6.3 & Error on RP\_GAIA & mag \\
PMRA\_GAIA & F7.3 & Gaia proper motion $\mu_{\rm \alpha}\cos{\rm \delta}$ & ${\rm mas\,yr^{-1}}$ \\
e\_PMRA\_GAIA & F7.3 & Error on PMRA\_GAIA & ${\rm mas\,yr^{-1}}$ \\
PMDEC\_GAIA & F7.3 & Gaia proper motion $\mu_{\rm \delta}$ & ${\rm mas\,yr^{-1}}$ \\
e\_PMDEC\_GAIA & F7.3 & Error on PMDEC\_GAIA & ${\rm mas\,yr^{-1}}$ \\
RV\_SDSS & F5.1 & Radial velocity SDSS & ${\rm km\,s^{-1}}$ \\
e\_RV\_SDSS & F5.1 & Error on RV\_SDSS & ${\rm km\,s^{-1}}$ \\
RV\_LAMOST & F5.1 & Radial velocity LAMOST & ${\rm km\,s^{-1}}$ \\
e\_RV\_LAMOST & F5.1 & Error on RV\_LAMOST & ${\rm km\,s^{-1}}$ \\
TEFF & F8.1 & Effective temperature & K \\
e\_TEFF & F8.1 & Error on T\_EFF & K \\
LOG\_G & F4.2 & Log surface gravity (gravity in ${\rm cm\,s^{-2}}$) & dex \\
e\_LOG\_G & F.4.2 & Error on LOG\_G & dex \\
LOG\_Y & F5.2 & Log helium abundance $n({\rm He})/n({\rm H})$ & dex \\
e\_LOG\_Y & F5.2 & Error on LOG\_Y & dex \\
PARAMS\_REF & A20 & Reference for atmospheric parameters (Bibcode) &  \\
EB-V & F6.4 & Instellar reddening E(B-V) & mag \\
e\_EB-V & F6.4 & Error on EB-V & mag \\
AV & F6.4 & Interstellar extinction A$_{\rm V}$ & mag \\
FUV\_GALEX & F6.3 & GALEX FUV-band magnitude & mag \\
e\_FUV\_GALEX & F6.3 & Error on FUV\_GALEX & mag \\
NUV\_GALEX & F6.3 & GALEX NUV-band magnitude & mag \\
e\_NUV\_GALEX & F6.3 & Error on NUV\_GALEX & mag \\
V\_APASS & F6.3 & APASS V-band magnitude & mag \\
e\_V\_APASS & F6.3 & Error on V\_APASS & mag \\
B\_APASS & F6.3 & APASS B-band magnitude & mag \\
e\_B\_APASS & F6.3 & Error on V\_APASS & mag \\
g\_APASS & F6.3 & APASS g-band magnitude & mag \\
e\_g\_APASS & F6.3 & Error on g\_APASS & mag \\
r\_APASS & F6.3 & APASS r-band magnitude & mag \\
e\_r\_APASS & F6.3 & Error on r\_APASS & mag \\
i\_APASS & F6.3 & APASS i-band magnitude & mag \\
e\_i\_APASS & F6.3 & Error on i\_APASS & mag \\
u\_SDSS & F6.3 & SDSS u-band magnitude & mag \\
e\_u\_SDSS & F6.3 & Error on u\_SDSS & mag \\
g\_SDSS & F6.3 & SDSS g-band magnitude & mag \\
e\_g\_SDSS & F6.3 & Error on g\_SDSS & mag \\
r\_SDSS & F6.3 & SDSS r-band magnitude & mag \\
e\_r\_SDSS & F6.3 & Error on r\_SDSS & mag \\
i\_SDSS & F6.3 & SDSS i-band magnitude & mag \\
e\_i\_SDSS & F6.3 & Error on i\_SDSS & mag \\
z\_SDSS & F6.3 & SDSS z-band magnitude & mag \\
e\_z\_SDSS & F6.3 & Error on z\_SDSS & mag \\
u\_VST & F6.3 & VST surveys (ATLAS, KiDS) u-band magnitude & mag \\
e\_u\_VST & F6.3 & Error on u\_VST & mag \\
g\_VST & F6.3 & VST surveys (ATLAS, KiDS) g-band magnitude & mag \\
e\_g\_VST & F6.3 & Error on g\_VST & mag \\
r\_VST & F6.3 & VST surveys (ATLAS, KiDS) r-band magnitude & mag \\
e\_r\_VST & F6.3 & Error on r\_VST & mag \\
i\_VST & F6.3 & VST surveys (ATLAS, KiDS) i-band magnitude & mag \\
e\_i\_VST & F6.3 & Error on i\_VST & mag \\
z\_VST & F6.3 & VST surveys (ATLAS, KiDS) z-band magnitude & mag \\
e\_z\_VST & F6.3 & Error on z\_VST & mag \\
u\_SKYM & F6.3 & SkyMapper u-band magnitude & mag \\
e\_u\_SKYM & F6.3 & Error on u\_SKYM & mag \\
v\_SKYM & F6.3 & SkyMapper v-band magnitude & mag \\
e\_v\_SKYM & F6.3 & Error on v\_SKYM & mag \\
g\_SKYM & F6.3 & SkyMapper g-band magnitude & mag \\
e\_g\_SKYM & F6.3 & Error on g\_SKYM & mag \\
r\_SKYM & F6.3 & SkyMapper r-band magnitude & mag \\
e\_r\_SKYM & F6.3 & Error on r\_SKYM & mag \\
i\_SKYM & F6.3 & SkyMapper i-band magnitude & mag \\
e\_i\_SKYM & F6.3 & Error on i\_SKYM & mag \\
z\_SKYM & F6.3 & SkyMapper z-band magnitude & mag \\
e\_z\_SKYM & F6.3 & Error on z\_SKYM & mag \\
g\_PS1 & F7.4 & PS1 g-band magnitude & mag \\
e\_g\_PS1 & F7.4 & Error on g\_PS1 & mag \\
r\_PS1 & F7.4 & PS1 r-band magnitude & mag \\
e\_r\_PS1 & F7.4 & Error on r\_PS1 & mag \\
i\_PS1 & F7.4 & PS1 i-band magnitude & mag \\
e\_i\_PS1 & F7.4 & Error on i\_PS1 & mag \\
z\_PS1 & F7.4 & PS1 z-band magnitude & mag \\
e\_z\_PS1 & F7.4 & Error on z\_PS1 & mag \\
y\_PS1 & F7.4 & PS1 y-band magnitude & mag \\
e\_y\_PS1 & F7.4 & Error on y\_PS1 & mag \\
J\_2MASS & F6.3 & 2MASS J-band magnitude & mag \\
e\_J\_2MASS & F6.3 & Error on J\_2MASS & mag \\
H\_2MASS & F6.3 & 2MASS H-band magnitude & mag \\
e\_H\_2MASS & F6.3 & Error on H\_2MASS & mag \\
K\_2MASS & F6.3 & 2MASS K-band magnitude & mag \\
e\_K\_2MASS & F6.3 & Error on K\_2MASS & mag \\
Y\_UKIDSS & F6.3 & UKIDSS Y-band magnitude & mag \\
e\_Y\_UKIDSS & F6.3 & Error on Y\_UKIDSS & mag \\
J\_UKIDSS & F6.3 & UKIDSS J-band magnitude & mag \\
e\_J\_UKIDSS & F6.3 & Error on J\_UKIDSS & mag \\
H\_UKIDSS & F6.3 & UKIDSS H-band magnitude & mag \\
e\_H\_UKIDSS & F6.3 & Error on H\_UKIDSS & mag \\
K\_UKIDSS & F6.3 & UKIDSS K-band magnitude & mag \\
e\_K\_UKIDSS & F6.3 & Error on K\_UKIDSS & mag \\
Z\_VISTA & F6.3 & VISTA surveys (VHS, VIKING) Z-band magnitude & mag \\
e\_Z\_VISTA & F6.3 & Error on Z\_VISTA & mag \\
Y\_VISTA & F6.3 & VISTA surveys (VHS, VIKING) Y-band magnitude & mag \\
e\_Y\_VISTA & F6.3 & Error on Y\_VISTA & mag \\
J\_VISTA & F6.3 & VISTA surveys (VHS, VIKING) J-band magnitude & mag \\
e\_J\_VISTA & F6.3 & Error on J\_VISTA & mag \\
H\_VISTA & F6.3 & VISTA surveys (VHS, VIKING) H-band magnitude & mag \\
e\_H\_VISTA & F6.3 & Error on H\_VISTA & mag \\
Ks\_VISTA & F6.3 & VISTA surveys (VHS, VIKING) Ks-band magnitude & mag \\
e\_Ks\_VISTA & F6.3 & Error on Ks\_VISTA & mag \\
W1 & F6.3 & WISE W1-band magnitude & mag \\
e\_W1 & F6.3 & Error on W1 & mag \\
W2 & F6.3 & WISE W2-band magnitude & mag \\
e\_W2 & F6.3 & Error on W2 & mag \\
W3 & F6.3 & WISE W3-band magnitude & mag \\
e\_W3 & F6.3 & Error on W3 & mag \\
W4 & F6.3 & WISE W4-band magnitude & mag \\
e\_W4 & F6.3 & Error on W4 & mag \\
\noalign{\smallskip}
\hline\hline
\end{longtable}

\end{document}